\documentclass[aps,preprint,groupedaddress]{revtex4}

\usepackage{graphics,epsfig}

\begin{document}

\title{Bounds for Lepton Flavor Violation and the Pseudoscalar Higgs in the General
Two Higgs Doublet Model using $g-2$ muon factor}
\author{Rodolfo A. Diaz}
\author{R. Martinez}
\author{J-Alexis Rodriguez}
\affiliation{Departamento de Fisica, Universidad Nacional de Colombia\\
Bogota, Colombia}

\begin{abstract}
Current experimental data from the $g-2$ muon factor, seems to
show the necessity of physics beyond the Standard Model (SM),
since the difference between SM and experimental predictions is
2.6$\sigma $. In the framework of the General Two Higgs Doublet
Model (2HDM), we calculate the muon anomalous magnetic moment to
get lower and upper bounds for the Flavour Changing (FC) Yukawa
couplings in the leptonic sector. We also obtain lower bounds for
the mass of the pseudoscalar Higgs ($m_{A^0}$) as a function of
the parameters of the model.
\end{abstract}

\maketitle

Current muon anomalous magnetic moment $a_{\mu }\;$ measurement has challenged Standard Model(SM)  and seems to open a window for new physics. Due to the high precision in $a_\mu$ measurement, it gives very restrictive bounds on physics beyond the SM.
Although $a_e$ measurement is about 350 more
precise \cite{marciano}, $a_{\mu }\;$is much more sensitive to New Physics
since contributions to $a_{l}\;$are
usually proportional to $m_{l}^{2}$.

The most accurate measurement of $a_{\mu }\;$hitherto, has been provided by
the Brookhaven Alternating Gradient Syncrotron \cite{g2 positive}. Their
data have an error one third that of the combined previous data \cite{Caso},
ref \cite{g2 positive} reports
\begin{equation}
a_{\mu ^{+}}=11659202\left( 14\right) \left( 6\right) \times 10^{-10}.
\label{g2 muon exp}
\end{equation}
On the other hand, SM predictions for $a_{\mu }$ has been estimated taking
into account the contributions from QED, Hadronic loops and electroweak
corrections. The final current result is \cite{marciano, g2 positive}
\begin{equation}
a_{\mu }^{SM}=11659159.6\left( 6.7\right) \times 10^{-10}.
\label{g2 muon SM}
\end{equation}

Taking into account (\ref{g2 muon exp}) and (\ref{g2 muon SM}) is obtained
\begin{equation}
\Delta a_{\mu }^{NP}=a_{\mu }^{\exp }-a_{\mu }^{SM}=42.6\left( 16.5\right)
\times 10^{-10},
\end{equation}
where $a_{\mu }^{\exp }\;$is the world average experimental value. Consequently at $90\%$
C.L.
\begin{equation}
21.5 \times 10^{-10}\leq \Delta a_{\mu }^{NP}\leq 63.7
\times 10^{-10}.   \label{Room for NP}
\end{equation}
$\Delta a_{\mu }^{NP}\;$gives the room available for New Physics, so $a_{\mu}^{exp}$ differs
from $a_{\mu}^{SM}$ approximately in $2.6\sigma $. Therefore, physics beyond the SM is needed to
achieve an acceptable theoretical experimental agreement. The most studied
contributions to $a_{\mu }\;$has been carried out in the framework of
radiative muon mass models as well as the Minimal Supersymmetric Standard
Model (MSSM), E$_{6}\;$string-inspired models, and extensions of MSSM with
an extra singlet \cite{estatus}.

Moreover, a very interesting suggestion to conciliate the new experimental
data with theoretical predictions is to consider models that includes
FCNC at tree level. Interactions involving FCNC are forbidden
at tree level in the SM, but could be present at one loop level as in the
case of $b\rightarrow s\gamma $ \cite{bsg}, $K^{0}\rightarrow \mu ^{+}\mu
^{-}$ \cite{kmm}, $K^{0}-\overline{K}^{0}$ \cite{koko}, $t\rightarrow
c\gamma $ \cite{tcg} etc. Many extensions of the SM permit FCNC at tree
level.\ For example, the introduction of new representations of fermions
different from doublets produce them by means of the Z-coupling \cite{2}.
Additionally, they are generated at tree level by adding a second doublet to
the SM \cite{wolf}, such couplings can be gotten as well in SUSY theories
without R-parity. Some other important new sources for FCNC might be
provided by a muon collider, as the processes $\mu \mu \rightarrow \mu \tau
(e\tau )\;$ mediated by Higgs exchange \cite{Workshop}, \cite{SherCollider},
which produce Lepton Flavor Violation (LFV).

However, there are several mechanisms to avoid FCNC at tree level. Glashow
and Weinberg \cite{gw} proposed a discrete symmetry to supress them in the
Two Higgs Doublet Model (2HDM) which is the simplest one that exhibits these
rare processes at tree level. There are two kinds of models which are
phenomenologically plausible with the discrete symmetry imposed. In the
model type I, one Higgs Doublet provides masses to the up-type and down-type
quarks, simultaneously. In the model type II, one Higgs doublet gives masses
to the up-type quarks and the other one to the down-type quarks. But the
discrete symmetry \cite{gw} is not compulsory and both doublets may generate
the masses of the quarks of up-type and down-type simultaneously, in such
case we are in the model type III \cite{III}. It has been used to search for
FCNC at tree level \cite{ARS}, \cite{Sher91}.

Recently, the 2HDM type III has been discussed and classified \cite{us},
depending on how the basis for the vacuum expectation values (VEV) are
chosen and according to the way in which the flavor mixing matrices are
rotated. In brief, the reference \cite{us} shows that there are two types of
rotations which generate four different lagrangians in the quark sector and
two different ones in the leptonic sector. The well known 2HDM types I and
II, could be generated from them in the limit in which the FC vertices
vanish. It has been pointed out that the phenomenology of the
2HDM type III is highly sensitive to the rotation used for the mixing
matrices.

In this paper, we calculate the
contributions to $\Delta a_{\mu }$ coming from the 2HDM,
which includes FCNC at tree level. We
will constrain the FC vertex involving the second and third
charged leptonic sector by using the result for $\Delta a_{\mu}^{NP}$, equation (\ref{Room for NP}). Additionally,
we get lower bounds on the pseudoscalar Higgs mass by taking into account
the lower experimental value of $\Delta a_{\mu }^{NP}\;$ at $90\%$ C.L.
by making reasonable assumptions on the FC vertex.

The Yukawa's Lagrangian for the 2HDM type III, is as follow
\begin{eqnarray}
-\pounds _{Y} &=&\eta _{ij}^{U}\overline{Q}_{iL}\widetilde{\Phi }%
_{1}U_{jR}+\eta _{ij}^{D}\overline{Q}_{iL}\Phi _{1}D_{jR}+\eta _{ij}^{E}%
\overline{l}_{iL}\Phi _{1}E_{jR}  \nonumber \\
&+&\xi _{ij}^{U}\overline{Q}_{iL}\widetilde{\Phi }_{2}U_{jR}+\xi _{ij}^{D}%
\overline{Q}_{iL}\Phi _{2}D_{jR}+\xi _{ij}^{E}\overline{l}_{iL}\Phi
_{2}E_{jR}+h.c.  \label{Yukawa}
\end{eqnarray}
where $\Phi _{1,2}\;$are the Higgs doublets,$\;\eta _{ij}\;$and $\xi _{ij}\;$%
are non-diagonal $3\times 3\;$matrices and $i$, $j$ are family indices.
In this work, we are interested only in neutral currents in the leptonic sector. We
also consider a CP-conserving model in which both Higgs doublets
acquire a VEV,
\begin{equation}
\left\langle \Phi _{1}\right\rangle _{0}=\left(
\begin{array}{c}
0 \\
v_{1}/\sqrt{2}
\end{array}
\right) \;\;,\;\;\left\langle \Phi _{2}\right\rangle _{0}=\left(
\begin{array}{c}
0 \\
v_{2}/\sqrt{2}
\end{array}
\right) . \nonumber
\end{equation}
The neutral mass eigenstates are given by \cite{moda}
\begin{eqnarray}
\left(
\begin{array}{c}
G_{Z}^{0} \\
A^{0}
\end{array}
\right) &=&\left(
\begin{array}{cc}
\cos \beta & \sin \beta \\
-\sin \beta & \cos \beta
\end{array}
\right) \left(
\begin{array}{c}
\sqrt{2}Im\phi _{1}^{0} \\
\sqrt{2}Im\phi _{2}^{0}
\end{array}
\right) ,  \nonumber \\
\left(
\begin{array}{c}
H^{0} \\
h^{0}
\end{array}
\right) &=&\left(
\begin{array}{cc}
\cos \alpha & \sin \alpha \\
-\sin \alpha & \cos \alpha
\end{array}
\right) \left(
\begin{array}{c}
\sqrt{2}Re\phi _{1}^{0}-v_{1} \\
\sqrt{2}Re\phi _{2}^{0}-v_{2}
\end{array}
\right)  \label{Autoestados masa Higgs}
\end{eqnarray}
where $\tan \beta =v_{2}/v_{1}\;$and $\alpha \;$is the mixing angle of the
CP-even neutral Higgs sector. $G_{Z}\;$is the would-be Goldstone boson of $Z$
and $A^{0}\;$is the CP-odd neutral Higgs.

Now, to convert the Lagrangian (\ref{Yukawa}) into mass eigenstates we make
the unitary transformations
\begin{equation}
E_{L,R}=\left( V_{L,R}\right) E_{L,R}^{0}\;\;  \label{Down transf}
\end{equation}
from which we obtain the mass matrices
\begin{equation}
M_{E}^{diag}=V_{L}\left[ \frac{v_{1}}{\sqrt{2}}\eta ^{E,0}+\frac{v_{2}}{%
\sqrt{2}}\xi ^{E,0}\right] V_{R}^{\dagger } \;\;  , \label{Masa down}
\end{equation}
where $M_E^{diag}$ is the diagonal mass matrix for the three lepton families.
From (\ref{Masa down}) we can solve for $\xi ^{E,0}\;$obtaining
\begin{equation}
\xi ^{E,0}=\frac{\sqrt{2}}{v_{2}}V_{L}^{\dagger }M_{E}^{diag}V_{R}-\frac{%
v_{1}}{v_{2}}\eta ^{E,0}.  \label{Rotation Id}
\end{equation}
which we call a rotation of type I. Replacing it into (%
\ref{Yukawa}), the expanded Lagrangian for the neutral leptonic sector is
\begin{eqnarray}
-\pounds _{Y\left( E\right) }^{\left( I\right) } &=&\frac{g}{2M_{W}\sin
\beta }\overline{E}M_{E}^{diag}E\left( \sin \alpha H^{0}+\cos \alpha
h^{0}\right)  \nonumber \\
&+&\frac{ig}{2M_{W}}\overline{E}M_{E}^{diag}\gamma _{5}EG^{0}+\frac{ig\cot
\beta }{2M_{W}}\overline{E}M_{E}^{diag}\gamma _{5}EA^{0}  \nonumber \\
&-&\frac{1}{\sqrt{2}\sin \beta }\overline{E}\eta ^{E}E\left[ \sin \left(
\alpha -\beta \right) H^{0}+\cos \left( \alpha -\beta \right) h^{0}\right]
\nonumber \\
&-&\frac{i}{\sqrt{2}\sin \beta }\overline{E}\eta ^{E}\gamma _{5}EA^{0}+h.c.
\label{Yukawa 1ad}
\end{eqnarray}
where the superindex $(I)\;$refers to the rotation type I.
It is easy to check that Lagrangian (\ref{Yukawa 1ad}) is just the one in
the 2HDM type I\ \cite{moda}, plus some FC interactions. Therefore, we
obtain the lagrangian of the 2HDM type I from eq (\ref{Yukawa 1ad}) by
setting $\eta ^{E}=0.\;$ In this case it is clear that when $\tan\beta\rightarrow 0$
then $\eta^E$ should go to zero, in order to have a finite contribution for FCNC at tree level.

On the other hand, from (\ref{Masa down}) we can also solve for $\eta
^{E,0}\;$instead of $\xi ^{E,0}$, to get
\begin{equation}
\eta ^{E,0}=\frac{\sqrt{2}}{v_{1}}V_{L}^{\dagger }M_{E}^{diag}V_{R}-\frac{%
v_{2}}{v_{1}}\xi ^{E,0}  \label{Rotation IId}
\end{equation}
which we call a rotation of type II. Replacing it into (\ref{Yukawa}) the
expanded Lagrangian for the neutral leptonic sector is
\begin{eqnarray}
-\pounds _{Y(E)}^{(II)} &=&\frac{g}{2M_{W}\cos \beta }\overline{E}%
M_{E}^{diag}E\left( \cos \alpha H^{0}-\sin \alpha h^{0}\right)  \nonumber \\
&+&\frac{ig}{2M_{W}}\overline{E}M_{E}^{diag}\gamma _{5}EG^{0}-\frac{ig\tan
\beta }{2M_{W}}\overline{E}M_{E}^{diag}\gamma _{5}EA^{0}  \nonumber \\
&+&\frac{1}{\sqrt{2}\cos \beta }\overline{E}\xi^{E}E\left[ \sin \left(
\alpha -\beta \right) H^{0}+\cos \left( \alpha -\beta \right) h^{0}\right]
\nonumber \\
&+&\frac{i}{\sqrt{2}\cos \beta }\overline{E}\xi ^{E}\gamma _{5}EA^{0} + h.c.
\label{Yukawa 2ad}
\end{eqnarray}
The Lagrangian (\ref{Yukawa 2ad}) coincides with
the one of the 2HDM type II \cite{moda}, plus some FC interactions. So, the
lagrangian of the 2HDM type II is obtained setting $\xi ^{E}=0.\;$ In this case it is clear
that when $\tan\beta\rightarrow \infty$
then $\xi^E$ should go to zero, in order to have a finite contribution for FCNC at tree level.

In the present report, we calculate $a_{\mu}$ in the 2HDM with FC interactions.
If we neglect the muon mass, the contribution at one loop
from all Higgses is given by
\begin{equation}
\Delta a_{\mu }^{NP} =\frac{m_{\mu }m_{l}}{16\pi ^{2}}\sum_{i}F\left(
m_{H_{i}},m_{l}\right)  a_{i}^2 \;,
\end{equation}
where
\begin{equation}
F\left( m_{H_{i}},m_{l}\right) =\allowbreak \frac{\widehat{m}%
_{H_{i}}^{2}\left( \widehat{m}_{H_{i}}^{2}-4\right) +\left[ 3+2\ln \left(
\widehat{m}_{H_{i}}^{2}\right) \right] }{m_{H_{i}}^{2}\left( 1-\widehat{m}%
_{H_{i}}^{2}\right) ^{3}}  \label{anomalo}
\end{equation}
$\allowbreak $ \allowbreak with $\widehat{m}_{H_{i}}=m_{l}/m_{H_{i}}$\ and $%
m_{l}\;$is the mass of the lepton running into the loop. The sum is over the
index $i=m_{h^{0}},m_{H^{0}},m_{A^{0}}$. The coefficients $%
a_{i}$ are the Feynman rules for FC couplings involved.

If we take into account the experimental data (\ref{Room for NP}), we get
some lower and upper bounds on the mixing vertex $\eta \left( \xi \right)
_{\mu \tau }\;$for the rotations of type I (II).
In figure 1, we display lower and upper bounds for the FC
vertices as a function of the $\tan\beta$ for both types of rotations with $m_{h^0}=m_{H^0}=150$
GeV and $m_{A^0}\rightarrow\infty$. For $\tan\beta=1$ the behaviour of the bounds for  both rotations is
the same. In the first case, rotation type I, the allowed region for
$\eta_{\mu\tau}$ is between $0.07\leq\eta_{\mu\tau}\leq 0.13$
for large values of $\tan\beta$. Meanwhile,
for rotation type II, the allowed region for small $\tan\beta$ is the same.
From Lagrangian (11), rotation type I, we can see that when
$\tan\beta\rightarrow 0$, $\eta_{\mu\tau}$ should go to zero as well to mantain a finite contribution to $\Delta a_\mu$.
This behaviour can be seen from figure 1. For rotation type II is similar to the former but in the limit
$\tan\beta\rightarrow \infty$.

In figure 2, we show lower and upper bounds for the FC vertex as a function of
$m_{H^0}$ for rotation of type II when $m_{h^{0}}=m_{H^{0}}$ and $m_{A^{0}}\rightarrow \infty $.
We see that the smaller value of $\xi_{\mu \tau }$ the larger value for
$\tan\beta$. We only consider the case of rotation type II because there is a
complementary behaviour as could be seen in figure 1.

Observe that according to the Feynman rules from (\ref{Yukawa 1ad}) and (\ref
{Yukawa 2ad}), the scalar (pseudoscalar) contribution to $\Delta a_{\mu
}^{NP}\;$eq. (\ref{anomalo}) is positive (negative). Such fact permits us to
impose lower bounds on the pseudoscalar Higgs mass, by using the lower limit in
eq. (\ref{Room for NP}). According to this equation the room for new physics
from $g-2$ muon factor is positive definite, and it is a new feature from most
updated results \cite{g2 positive}.

Now, to take into account the
experimental value (\ref{Room for NP}), we should make a supposition about
the value of the FC vertex. A reasonable assumption consists of taking
the geometric average of the Yukawa couplings \cite{Cheng Sher} i.e. $\eta
\left( \xi \right) _{\mu \tau }\approx 2.5\times 10^{-3}$. Additionally, we
shall use also the values $\eta \left( \xi \right) _{\mu \tau }\approx
2.5\times 10^{-2}$ and$\;\eta \left( \xi \right) _{\mu \tau }\approx
2.5\times 10^{-4}\;$which are one order of magnitude larger and smaller than
the former.  Using these suppositions and
the experimental value (\ref{Room for NP}) we get restrictions for $m_{A^{0}}\;$ and they
are plotted in figures (3)-(5).

Figure 3 displays $m_{A^{0}}\;$vs$\;\tan \beta \;$using rotation type II
with the three values of $\xi _{\mu \tau }\;$mentioned above and setting $%
m_{h^{0}}=m_{H^{0}}\;$with $m_{h^0}=110,\;300$ GeV. It could be seen that in the
limit of large $\tan \beta ,\;$the lower limit reduces to$\;m_{A^{0}}\approx
m_{h^0}$. The same behavior can be seen in rotation type I but the bound $%
m_{A^{0}}\approx m_{h^0}$ is gotten in the limit of small $\tan \beta .$ We see that
the smaller value of $\xi_{\mu \tau }\;$the stronger lower limit for $m_{A^{0}}$.

Figure 4 shows $m_{A^{0}}\;$vs$\;m_{h^{0}}\;$with $\xi_{\mu \tau }=2.5\times 10^{-3},
2.5\times 10^{-2}\;$and setting $\tan \beta
=1,\;\alpha =\pi /2\;$and using $m_{h^{0}}=m_{H^{0}}=110$ GeV. With
this settings, the value $\xi_{\mu \tau }= 2.5\times
10^{-4}\;$ is excluded. Using such specific arrangements, the bounds are
identical in both types of rotations.

In figure 5 we suppose that $m_{h^0}\neq m_{H^0}$ and the bounds are function of $\sin\alpha$.
 The above figure shows the sensitivity of lower
bounds on $m_{A^{0}}\;$ with the mixing angle $\alpha ,\;$for rotation type II.
The value $\xi _{\mu \tau }=2.5\times 10^{-4}\;$is excluded again. The
constraints are very sensitive for $\xi _{\mu \tau }=2.5\times 10^{-3}\;$ respect to
$\sin\alpha$ but
rather insensitive for $\xi _{\mu \tau }=2.5\times 10^{-2}$.
The figure below shows $m_{A^{0}}\;$vs\ $\tan \beta \;$for $m_{h^{0}}=110$ GeV,
$m_{H^{0}}=300$ GeV, $\alpha =\pi /6,\;$ for rotation type II and using the
same three values of $\xi _{\mu \tau }.\;$The $m_{A^0}$ lower asymptotic limit for large $\tan
\beta \;$is approximately $m_{h^{0}}$.

In conclusion, we have found lower and upper bounds for the FC vertex $%
\eta \left( \xi \right) _{\mu \tau }\;$in the context of the general 2HDM by using
the allowed range for $\Delta a_{\mu }^{NP}\;$ at $90\%$ C.L. and$\;$utilizing several sets
of values for parameters of the model. Additionally, in the limit $m_{A^0} \to \infty$ we get that for small (large)
values of $\tan \beta \;$the allowed range for the FC vertex $\eta _{\mu
\tau }\left( \xi _{\mu \tau }\right) \;$becomes narrower, and both upper and
lower bounds go to zero in the rotation of type I (II).

On the other hand, we have gotten lower bounds on the pseudoscalar Higgs
mass of the 2HDM coming from the $g-2\;$muon factor, by using the
experimental value of $\Delta a_{\mu }^{NP}\;$and making reasonable
assumptions on the FC vertex $\eta \left( \xi \right) _{\mu \tau }$.
Specifically, we have taken for $\eta \left( \xi \right) _{\mu \tau }$ the
geometric average of the Yukawa couplings, and we also utilized  values one
order of magnitude larger and one
smaller. Taking these three values for the FC vertex we find that the
smaller value for $\eta \left( \xi \right) _{\mu \tau }\;$the more stringent
lower bounds for $m_{A^{0}}\;$. Additionally, assuming $%
m_{H^{0}}=m_{h^{0}},\;$we show that in the limit of small (large) $\tan
\beta \;$the lower bound of $m_{A^{0}}\;$becomes merely $m_{A^{0}}\approx m_{h^0}\;$%
for rotation of type I (II). In the case of different scalar masses,
there is still a lower asymptotic limit for $m_{A^{0}}$.
Notwithstanding, these lower constraints on $m_{A^{0}}\;$should be consider
carefully, since for $\eta \left( \xi \right) _{\mu \tau }\;$we can only
make reasonable estimations but they are unknown so far.

This work was supported by COLCIENCIAS, DIB and DINAIN.

\begin{figure}[h]
\begin{center}
\includegraphics[angle=0, width=10cm]{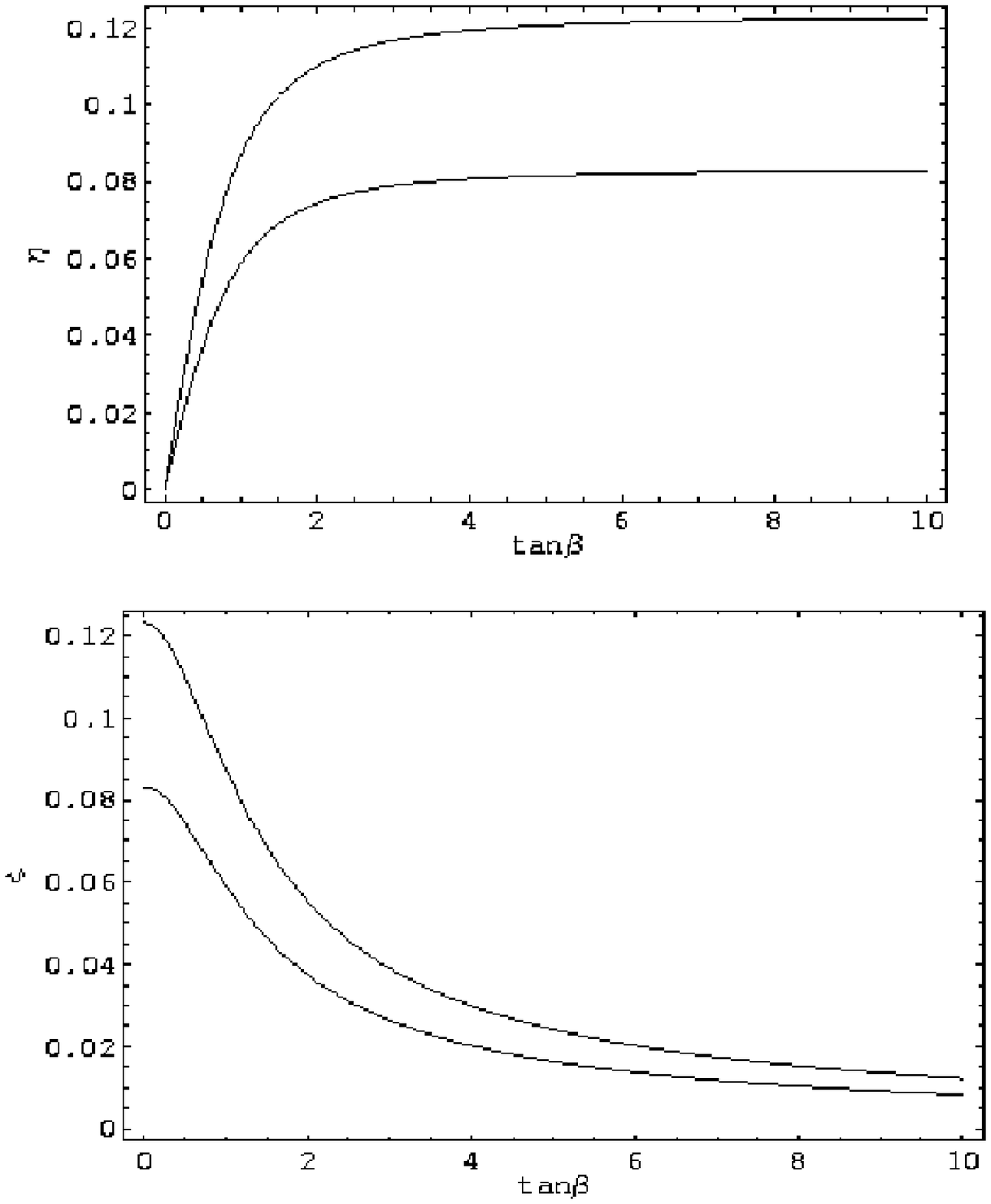}
\end{center}
\caption{ Lower and upper bounds for $\eta _{\mu \tau }\left( \xi _{\mu \tau}\right) \;$vs\
 tan$\beta ,\;$for rotations I and II using
$m_{h^{0}}=m_{H^{0}}=150$ GeV and $m_{A^{0}}\rightarrow \infty .\;$}
\label{Fig. 1}
\end{figure}
\newpage
\begin{figure}[h]
\begin{center}
\includegraphics[angle=0, width=10cm]{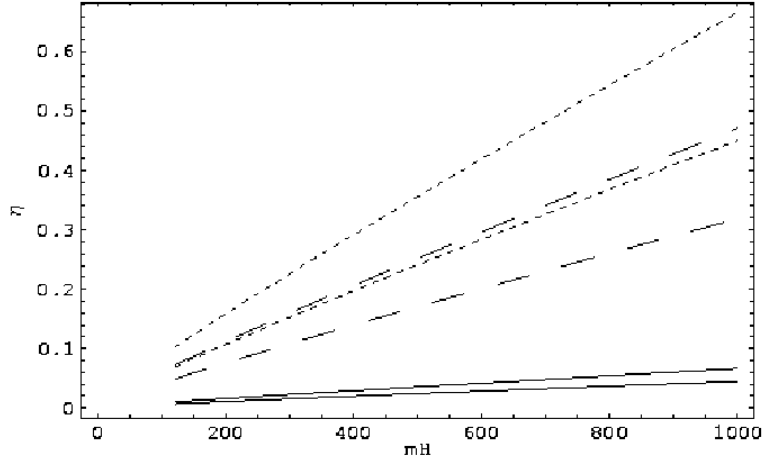}
\end{center}
\caption{Figure 2. Lower and upper bounds for $\eta _{\mu \tau }\left( \xi _{\mu \tau
}\right) \;$vs\  $m_{H^{0}},\;$for rotation of type I, taking $m_{h^{0}}=m_{H^{0}}\;$and $m_{A^{0}}\rightarrow \infty$, the pair of
short dashed lines correspond to $\tan \beta =30,\;$the long dashed lines
are for $\tan \beta =1$,\ and the continuous lines are for $\tan \beta =0.1.$}
\label{Fig. 2}
\end{figure}
\newpage
\begin{figure}[h]
\begin{center}
\includegraphics[angle=0, width=10cm]{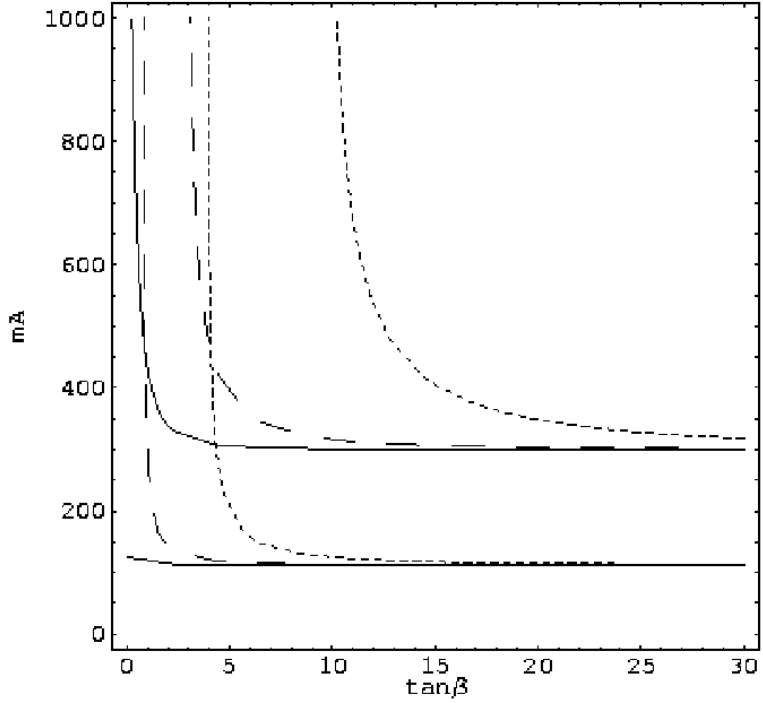}
\end{center}
\caption{ Contour plot of  $m_{A^{0}}\;$vs$\;\tan \beta \;$using rotation
type II and assuming $m_{h^{0}}=m_{H^{0}}$. Short dashed lines correspond to $\xi _{\mu
\tau }=2.5\times 10^{-4}\;$for $m_{H^0}=110\;$ GeV (below) and $m_{H^0}=300\;$ GeV (above). Long
dashed lines correspond to $\xi _{\mu \tau }=2.5\times 10^{-3}\;$for $m_{H^0}=110\;
$ GeV (below) and $m_{H^0}=300\;$ GeV (above). Finally, solid lines correspond to $\xi
_{\mu \tau }=2.5\times 10^{-3}\;$for $m_{H^0}=110\;$ GeV (below) and $m_{H^0}=300\;$ GeV (above).}
\label{Fig. 3}
\end{figure}
\newpage

\begin{figure}[h]
\begin{center}
\includegraphics[angle=0, width=10cm]{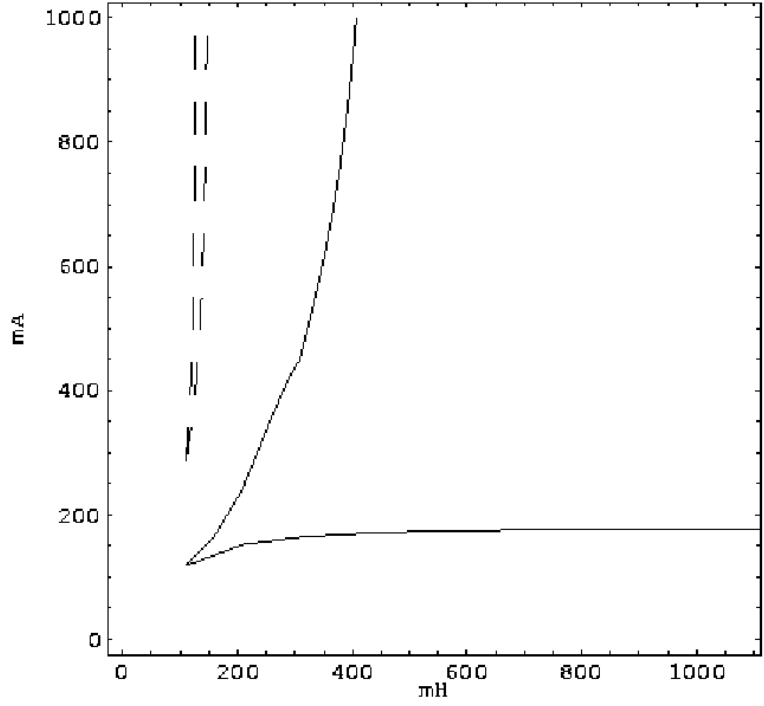}
\end{center}
\caption{ Contour plot of$\;m_{A^{0}}\;$vs$\;m_{h^{0}}\;$setting $\tan \beta
=1,\;\alpha =\pi /2.\;$Long dashed lines correspond to $\xi _{\mu \tau
}=2.5\times 10^{-3}\;$for $m_{H^{0}}=110\;$ GeV (below) and $m_{h^{0}}=m_{H^{0}}\;
$(above). Solid lines correspond to $\xi _{\mu \tau }=2.5\times
10^{-2}\;$for $m_{h^{0}}=110\;$ GeV (below) and $m_{h^{0}}=m_{H^{0}}\;$(above).}
\label{Fig. 4}
\end{figure}
\newpage
\begin{figure}[h]
\begin{center}
\includegraphics[angle=0, width=10cm]{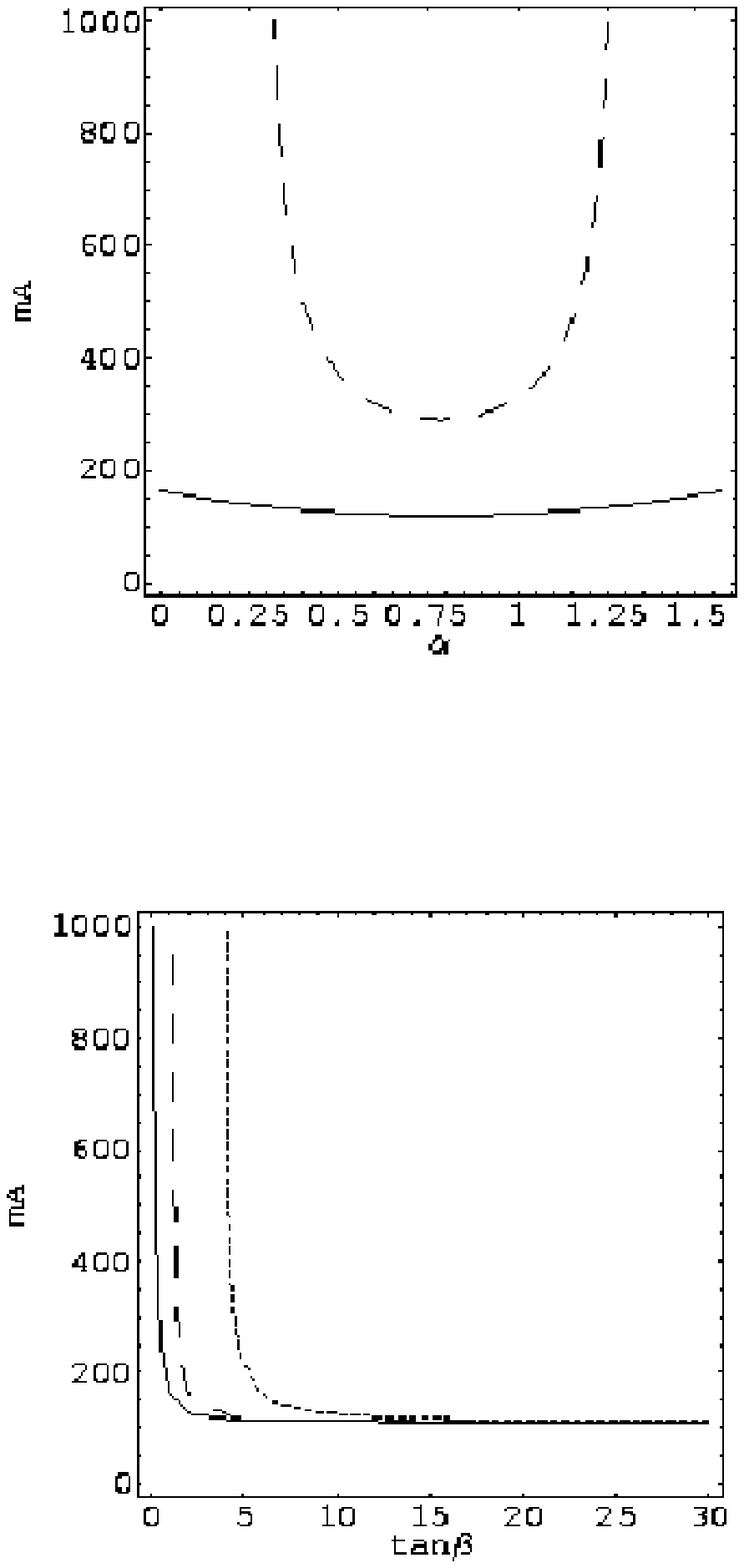}
\end{center}
\caption{(top) Contour plot of $m_{A^{0}}\;$vs\ $\alpha ,\;$for rotation type
II. Dashed line correspond to $\xi _{\mu \tau }=2.5\times 10^{-3}$ ,\
solid line correspond to $\xi _{\mu \tau }=2.5\times 10^{-2}$.
(bottom) Contour plot of $m_{A^{0}}\;$vs\ $\tan \beta \;$for
$m_{H^{0}}=300$ GeV, $m_{h^{0}}=110$ GeV, $\alpha =\pi /6,\;$and for rotation type II.
Short dashed line correspond to $\xi _{\mu \tau }=2.5\times 10^{-4}$, long
dashed line correspond to $\xi _{\mu \tau }=2.5\times 10^{-3}$, and
solid line correspond to $\xi _{\mu \tau }=2.5\times 10^{-2}.\;$}
\label{Fig. 5}
\end{figure}

\end{document}